\documentclass[a4paper,12pt]{amsart}     

\usepackage{graphicx}
\usepackage{mathptmx}

\usepackage[latin1]{inputenc}

\usepackage{amsfonts}
\usepackage{amsmath}
\usepackage{amssymb}
\usepackage[all]{xy}
\usepackage{graphicx}
\usepackage{amscd}

\newtheorem{theorem}{Theorem}[section]
\theoremstyle{definition}
\newtheorem{definition}[theorem]{Definition}
\theoremstyle{remark}
\newtheorem{remark}[theorem]{Remark}

\newcommand{\F}{\mathbb{F}}

\newcommand{\C}{\mathcal{C}}
\newcommand{\Ocal}{\mathcal{O}}

\newcommand{\Lcal}{\mathcal L}
\newcommand{\Ical}{\mathcal I}
\newcommand{\Pmat}{\mathbb{P}}
\newcommand{\Amat}{\mathbb{A}}
\newcommand\Proj{\operatorname {Proj}}
\newcommand{\df}{d_{free}}
\newcommand{\w}{\mbox{${\rm w}$}}

\begin{document}

\title[Convolutional Goppa codes on fibrations]{Convolutional Goppa codes defined on fibrations}

\author[J. I. Iglesias Curto et al.]{J. I. Iglesias Curto,  J. M. Mu\~{n}oz Porras, F. J. Plaza Mart\'{\i}n \\ and
    G. Serrano Sotelo}

\thanks{This research was supported by the Spanish DGESYC
through research project MTM2009-11393 MICINN and by the ``Junta
de Castilla y Le\'on'' through research project SA029A08.
}



\address{
              Departamento de Matem\'{a}ticas, Universidad de Salamanca, Plaza de la Merced 1, 37008 Salamanca, Spain \\
              Tel.: +34+923294500 ext. 1553\\
              Fax: +34+923294583\\
}
              \email{joseig@usal.es}           
              \email{jmp@usal.es}           
              \email{fplaza@usal.es}           
              \email{laina@usal.es}           


\maketitle

\begin{abstract}
We define a new class of Convolutional Codes in terms of fibrations
of algebraic varieties generalizaing our previous constructions of
Convolutional Goppa Codes (\cite{DMS:04,MDIS:06}). Using this general
construction we can give several examples of Maximum Distance
Separable (MDS) Convolutional Codes.
\end{abstract}

\section{Introduction}

This paper offers a generalization of our algebro-geometric construction
of Convolutional Goppa Codes (CGC) (\cite{DMS:04,DMS:08,JIIC:09,JIIC:09b,MDIS:06}). Recall, that Algebraic Geometry has been successfully
applied in Coding Theory during the last decades, first for block codes
(e.g. \cite{Gop:77,Gop:81,Pir88,vLvG,HovLP}) and more recently for
convolutional codes (\cite{Lom:01,RR:94}), as a natural continuation of
the algebraic constructions of convolutional codes already known for long
(as for example \cite{Jus:75,Pir:76}). Our aim is to consider algebro-geometric properties of higher
dimensional varieties in order to obtain Convolutional Goppa Codes with good properties, as it
has been successfully done for block codes (\cite{Han:99,VZ:05}).

CGC are constructed in terms of families of algebraic varieties,
$X\to {\mathbb A}^1$, parametrized by an affine line ${\mathbb
A}^1$. In the case of block codes, the generalization of Goppa Codes
to higher-dimensional varieties has been successfully used.

The contents of this work are arranged in the following way. In
\S\ref{sec:prelim} we summarize some notions and results on
convolutional codes based on \cite{For:70,McE:98}. In
\S\ref{sec:generalconstruction} we expose the construction of CGC
defined by a family of algebraic varieties, $X\to {\mathbb A}^1$,
parametrized by the affine line. This construction consists of
evaluating sections of an invertible sheaf on sections of the
fibration $X\to {\mathbb A}^1$. In \S\ref{sec:CodesP2General}
details are given for the construction when consider in the
particular case of the trivial fibration $\Pmat^2_{\F_q}\times
\Amat^1 \to \Amat^1$. \S\ref{sec:examples} provides some examples in order to
illustrate the possibilities of our approach.

We use the standard notations of Algebraic Geometry as can be found in \cite{Hart:77}.

\section{Preliminaries on Convolutional Codes}\label{sec:prelim}

Let $\F_q$ be a finite field of size $q=p^s$, with $p$ a prime.

Recall that, opposed to the definition of block codes as
vector subspaces, convolutional codes are (roughly) defined as
submodules of $\F[z]^n$. Convolutional codewords are then polynomial
vectors; indeed, the encoded sequence
$(c_0,c_1,c_2,\ldots)$ (with $c_i\in
\F^n$)  is equivalently represented as the polynomial  $c(z)=\sum_{i=0} c_i z^i $. Each codeword (encoded sequence) results of
applying a polynomial generator matrix
to an information word (or information sequence) $(u_0,u_1,u_2,\ldots)$, which will be analogously written as a polynomial
$u(z)=\sum_{i=0} u_i z^i$. The fact that the entries of the
generator matrix are polynomials implies that each encoded block
$c_i$ depends not only on $u_i$ but also on the previous information
blocks $u_{i-1},\ldots$. This is the distinctive aspect between
block and convolutional codes. The term ``convolutional'' is used
since the output sequences can be regarded as the convolution of the
input sequences with the sequences in the encoder, \cite{For:70}.
The control matrix (a.k.a. parity check matrix) and dual code for convolutional codes are
defined exactly in the same way as for block codes.

More rigourously, an $(n,k)$ \emph{convolutional code} $\C$ over $\F_q$ is defined as a rank $k$ submodule of  $\F_q[z]^n$. The integers $(n,k)$ are called, respectively, the \emph{length} and \emph{dimension} of the convolutional code. The quotient $\frac{k}{n}$ is called the \emph{rate} of the code.

Every $k\times n$ matrix of maximal rank, $G$, with entries in $\F_q[z]$ defines an injective map
    $$
    G\colon F[z]^k \hookrightarrow F[z]^n
    $$
Its image defines a $(n,k)$ convolutional code $\C$  and, in this case, $G$ will be called a \emph{polynomial encoder} or \emph{generator matrix} of $\C$, although not every generator matrix is equally suitable. We will prefer matrices which are basic; we say that  $G$ is \emph{basic} if the g.c.d. of the minors of order $k$ of $G$ is equal to $1$.

Any polynomial encoder $G$  for $\C$ induces a injective $\F_q(z)$-linear map
$$
G\colon  \F_q(z)^k\hookrightarrow \F_q(z)^n\, .
$$

Although this may allow us to generalize the notion of convolutional
code as subspaces of $\F_q(z)^n$, one has to bear in mind that
different associated encoders $G$ generate submodules of $\F_q[z]^n$
which may be different. However, basic encoders always generate the
same submodule (\cite{DMS:08}). Therefore, in this sense we may consider that the
notions of convolutional codes as submodules of $\F_q[z]^n$ or as
vector subspaces of $\F_q(z)^n$ are equivalent.

Indeed, for a given convolutional code $\C$, the unimodular group
$GL(k,\F_q[z])$ acts transitively on the set of basic encoders for
$\C$ (\cite{DMS:08}). Then, one can consider an
invariant associated with the code, the {\emph{ degree}} of the
code, $\delta$, defined as (e.g.
\cite{McE:98})
    $$
    \delta\,:=\, \text{maximum degree of the minors of order $k$ of a basic encoder for $\C$}
    $$

The \emph{degree} of a polynomial encoder $G$, $\deg G$, is the sum
of the degrees of its rows. Forney (\cite{For:70}) proved that for
each $(n,k)$-convolutional code of degree $\delta$ there exists at
least one basic encoder $G$ such that
    $$\delta=\deg G\leq \deg G'\,,$$
for all polynomial encoders $G'$ of the convolutional code. These
basic encoders $G$ are called \emph{minimal basic encoders} by
Forney \cite{For:70} or \emph{canonical encoders} by McEliece
\cite{McE:98}.

As for block codes, there is a notion of distance that will characterize the
error detection/correction capacity of convolutional codes the \emph{free
distance}, $\df$. Let us define the overall Hamming weight of a
polynomial vector $v(z)=\sum_{i=0} v_i z^i$ as $\w(v(z))=\sum_{i=0}
\w(v_i)$. Then the free distance of the code $\C$ is defined as
\begin{equation}
    d_{free}(\C)\,:=\, \underset{c\in \C}{\rm min\,} \w(c) \,.
\end{equation}
The free distance is directly related to the other parameters of the
code. However, the exact relationship is not known and different
bounds are considered instead. One of the most usually considered
is the generalized Singleton bound \cite{RS:99}
    \begin{equation}\label{eq:SingletonBound}
    \df\leq S(n,k,\delta)= (n-k)\left(\left\lfloor\frac{\delta}{k}
    \right\rfloor +1\right) +\delta +1
    \end{equation}
Convolutional codes attaining the generalized Singleton bound are
called Maximum Distance Separable (MDS).

Convolutional codes do have one more parameter which do not
have a counterpart in block codes, the \emph{memory}. It is well known, e.g. \cite{McE:98}, that the row
degrees of a canonical generator matrix are, up to ordering,
uniquely determined by the code. They are known as the \emph{Forney
indices} of the code. The largest of them is called the memory of
the code and denoted $m$. The sum of the Forney indices, which is
equal the maximal degree of the minors of any basic matrix, coincides the
degree (also known as complexity) of the code.
Roughly, the degree of the code measures the dependance of an
encoded block with respect to the information blocks, while the
memory counts on how many information blocks does every encoded
block depend. Convolutional codes of degree 0 are precisely linear
block codes.\\

\section{General construction}\label{sec:generalconstruction}

Let
$X$ be a variety of dimension $m+1\geq 2$, let $\Amat^1={\rm Spec\,}
{\F_q}[z]$ denote the affine line and let us consider a flat and
projective morphism $\pi:X\to \Amat^1 $ whose fibers  are smooth and geometrically irreducible algebraic varieties of dimension $m$. Recall that for ${\rm dim\,}X=2$  the fibers are curves; this case has been studied in \cite{DMS:04,MDIS:06}. For the basic facts on algebraic geometry that will be used here, we address the reader to~\cite{Hart:77}.

Let us choose $n$ different sections of $\pi$
$$
    p_i: \Amat^1\to X \qquad \text{with }p_i\circ\pi=\operatorname{Id} \quad \forall i=1,\ldots,n
$$
and, thus,  $p_i(\Amat^1)\subset X$ is a  curve isomorphic to
$\Amat^1$. Consider the closed subscheme
$$
    D=p_1(\Amat^1)\cup\ldots\cup p_n(\Amat^1)
$$
as well as the morphism $p$ given by the composition
$$
\xymatrix{
    D \ar@{^(->}[r]\ar[rd]^p & X \ar[d]^\pi \\
     & \Amat^1
}
$$
which is flat and finite of degree $n$.

Let us call $\Ocal_D$ and $\Ical_D$ respectively the sheaves of
rings and ideals of $D\hookrightarrow X$. We have an exact sequence
$$
    0\to \Ical_D\to \Ocal_X
        \to \Ocal_D \to 0
$$
Let $\Lcal$ be an
invertible sheaf over $X$. The tensor product of the sequence with
$\Lcal$ yields
$$
    0\to \Lcal\otimes\Ical_D\to \Lcal  \to \widetilde{\Ocal_D} \to 0\,,
$$
where $\widetilde{\Ocal_D}= \Ocal_D\underset{\Ocal_X}\otimes\Lcal$.
Taking global sections we obtain the long exact sequence of
$\F_q[z]$-modules
    {\small \begin{equation}\label{eq:exactsequence}
    \xymatrix@C=12pt{
    0 \ar[r] & H^0(X,\Lcal\otimes\Ical_D) \ar[r] &
    H^0(X,\Lcal) \ar[r] &
    H^0(X,\widetilde{\Ocal_D}) \ar[r] &
    H^1(X,\Lcal\otimes\Ical_D) \ar[r] &
    H^1(X,\Lcal) \ar[r] & 0}
    \end{equation}}

\begin{remark}
Note that the flatness of $p:D\to \Amat^1$ implies the existence of
isomorphisms $\phi:p_*\Ocal_D \overset{\sim}{\to}{\F_q}[z]^n$, where
$\F_q[z]$ also denotes its corresponding sheaf on $\Amat^1$. In
general these isomorphisms are not canonical but, if the chosen
sections $p_i$ are disjoint, then there exists a canonical
isomorphism $p_*\Ocal_D\simeq {\F_q}[z]^n$ induced by $p$.\\
\end{remark}

Since $\Lcal$ restricted to the sections is trivial, then
$\widetilde{\Ocal_D} \simeq \Ocal_D$, but such identification is not
canonical. Thus, if we fix an isomorphism
$$
    \phi: H^0(X,\Ocal_D)\overset{\sim}{\longrightarrow} {\F_q}[z]^n \,,
$$
as well as trivializations for each section $p_i$, we obtain induced
isomorphisms $\widetilde{\Ocal_D}\simeq\nolinebreak{}\Ocal_D$ and
$\psi: H^0(X,\widetilde{\Ocal_D}) \overset{\sim}{\to} {\F_q}[z]^n$.

\begin{remark}\label{remark:OXH}
If we take $\Lcal\simeq \Ocal_X(H)$, being $H$ an effective divisor
on $X$ which is flat over $\Amat^1$, the trivializations
$\Lcal|_{p_i}$ and the isomorphism $\psi$ are fixed.
\end{remark}


\begin{definition}\label{def:definitionGoppa}
The \emph{convolutional Goppa code} $\C(\Gamma,D,\psi)$ determined by the sheaf
$\Lcal$, the subscheme $D$, the isomorphism $\psi$,
 and a submodule $\Gamma\subseteq H^0(X,\Lcal)$  is the submodule given by the image of  the homomorphism $f$ defined by
$$
\xymatrix{
    0 \ar[r] & H^0(X,\Lcal\otimes\Ical_D) \ar[r] &
        H^0(X,\Lcal) \ar[r] &
        H^0(X,\widetilde{\Ocal_D}) \ar[d]^\wr_\psi \\
     & & \Gamma \ar@{^(->}[u] \ar@{-->}[r]^f & {\F_q}[z]^n
}
$$
\end{definition}

By the very construction,  the \emph{length of the code} is given by
the rank of $\Ocal_D$ as an $\Ocal_{\Amat^1}$-module, which is the
number $n$ of sections taken to define the code. The issue of
constructing such sections is clearly related with the question of
finding rational points in algebraic varieties over finite fileds.

The  \emph{dimension of the code} is equal to the rank of the
submodule $\operatorname{Im}(f)$ which coincides with the rank of $\Gamma$ if and only if $\Gamma\cap H^0(X,\Lcal\otimes\Ical_D)=(0)$. Note that in the case of the complete linear series, $\Gamma = H^0(X,\Lcal)$, the additive property of the dimension applied to the exact sequence \ref{eq:exactsequence} yields
$$
   \operatorname{rk} \C(\Gamma,D,\psi)=
    h^0(\Lcal)-h^0(\Lcal\otimes \Ical_D) =
    h^0(\widetilde{\Ocal_D})-h^1(\Lcal\otimes \Ical_D)+h^1(\Lcal)
$$
Nevertheless, the explicit calculus of these numbers for the general case is a
very hard problem in classical algebraic geometry based on the
theory of syzygies.

For the approach in terms of subspaces of ${\F_q}(z)$, one considers the the generic point of $\Amat^1$, $\eta$, whose residue field
is ${\F_q}(\eta)={\F_q}(z)$. The fiber $X_\eta$ is an
$m$-dimensional variety over ${\F_q}(z)$, and
$p_1(\eta),\ldots,p_n(\eta)$ are $n$ different ${\F_q}(z)$-rational
points. Then we have $D_\eta=p_1(\eta)\cup\ldots\cup p_n(\eta)$, and
an isomorphism
    $$
    \psi_\eta: H^0(X_\eta,\widetilde{\Ocal_{D_\eta}})
        \overset{\sim}{\to} {\F_q}(z)^n \, .
    $$
Moreover, if $\Lcal=\Ocal_X(H)$ then $\psi_\eta$ is canonical.

\begin{definition}\label{def:definitionGoppaSubspace}
The \emph{convolutional Goppa code}
$\C(\Gamma,D_\eta,\psi_\eta)$ is the image of the homomorphism
$f_\eta$ defined by
$$
\xymatrix{
    0 \ar[r] & H^0(X_\eta,\Lcal_\eta\otimes\Ical_{D_\eta}) \ar[r] &
        H^0(X_\eta,\Lcal_\eta) \ar[r] &
        H^0(X_\eta,\widetilde{\Ocal_{D_\eta}}) \ar[d]^\wr_{\psi_\eta} \\
     & & \Gamma \ar@{^(->}[u] \ar@{-->}[r]^{f_{\eta}}  & {\F_q}(z)^n
}
$$
where $\Gamma$ is a given subspace of $H^0(X_\eta,\Lcal_\eta)$
\end{definition}

The length and dimension of the code
$\C(\Gamma,D_\eta,\psi_\eta)$ are computed as above.

In the rest of the paper we will continue with the submodule
approach, but as we have just seen the shift to the subspace setting
would be straightforward.

As it was already mentioned in the Introduction, the case when ${\rm
dim\,}X=2$ has been already studied in \cite{DMS:04,MDIS:06}, in
particular for $X=\Pmat^1\times \Amat^1$. In the next Section we will
illustrate in detail how the construction works for higher
dimensional varieties by considering $X=\Pmat^2_{\F_q}\times \Amat^1$.

\section{Codes defined on the projective plane over $\F_q$}\label{sec:CodesP2General}

Let $\Pmat^2_{\F_q}= \Proj \F_q[x_0,x_1,x_2]$ be the projective
plane over $\F_q$, and let
    $$
    X=\Pmat^2_{\F_q}\times \Amat^1 \overset{\pi}\to \Amat^1 \;.
    $$
be the trivial fibration.

Let $H_\infty\subset\Pmat^2_{\F_q} $ be the line defined by the equation $x_0=0$. Then, its complement is an affine plane,  $\Pmat^2_{\F_q}\setminus H_\infty = \Amat^2$.

We will choose the sections $p_i$ ($1\leq i\leq n$) of $\pi$ taking
values in $ \Amat^2 \times \Amat^1$. They are given by
  \begin{align*}
    \Amat^1 & \xrightarrow{p_i} \Amat^2 \times \Amat^1 \\
    z &\mapsto p_i(z)=(\alpha_{i,1}z+\beta_{i,1},\alpha_{i,2}z+\beta_{i,2},z)
    \end{align*}
where all $\alpha_{i,r},\beta_{i,s}\in{\F_q}$. Observe that the length of the code is bounded by the number of different sections; that is, $n\leq q^4$. However, this count includes also linear codes (e.g. $\alpha_{i,r}\equiv 0$ for all $i,r$) as well as codes defined with the fibration $\Pmat^1_{\F_q}\times \Amat^1 \overset{\pi}\to \Amat^1$ (e.g. when the $n$ sections are collinear).

Let us considerer the divisor $\pi_1^*H_\infty$, where $\pi_1\colon X\to\Pmat^2_{\F_q}$ is the projection onto the first factor, and the
invertible sheaf given by
    $$\Lcal:=
    \Ocal_{\Pmat^2_{\F_q[z]}}(\pi_1^*H_\infty)^{\otimes r}\simeq
    \pi^*_1 \Ocal(1)^{\otimes r}=
    \Ocal(r)\underset{{\F_q}}\otimes {\F_q}[z]
    $$
 where we write $\Ocal(r):= \Ocal_{\Pmat^2_{\F_q}}(H_\infty)^{\otimes r}$ for simplicity.

Notice that, as it was pointed out in Remark \ref{remark:OXH}, the trivializations of
$\Lcal$ on the sections $p_i$ and the isomorphism $\psi: H^0(X,\widetilde{\Ocal_D})
\overset{\sim}{\to} {\F_q}[z]^n$ are fixed.

If we denote $t=\frac{x_1}{x_0}$, $s=\frac{x_2}{x_0}$ the affine coordinates in the affine plane $\Amat^2$,  the space of global sections is explicitly described as
$$
    H^0(\Pmat^2_{\F_q},\Ocal(r))\,=\,
    \langle t^i s^j \; \vert\; 0\leq i+j\leq r\rangle
$$
Hence, the evaluation of $ t^i s^j$ at the points $p_1,\ldots, p_n$ is given by
$$
    f( t^i s^j) =
        ( (\alpha_{1,1}z+\beta_{1,1})^i\cdot
        (\alpha_{1,2}z+\beta_{1,2})^j,\dots,
        (\alpha_{n,1}z+\beta_{n,1})^i\cdot
        (\alpha_{n,2}z+\beta_{n,2})^j ).
$$

Since $H^1(X,\Lcal)=0$ and $H^0(X,\Lcal)= H^0(\Pmat^2_{\F_q} , \Ocal(r))\otimes_{\F_q} \F_q[z]$, the long exact sequence~(\ref{eq:exactsequence}) reads now as follows
    {\small
    $$
    \xymatrix@R8pt@C14pt {
     0 \ar[r] &  H^0(X,\Lcal\otimes\Ical_{D}) \ar[r] &
     H^0(\Pmat^2_{\F_q} , \Ocal(r))\otimes_{\F_q} \F_q[z]  \ar[r] &
     \F_q[z]^n \ar[r] &
     H^1(X,\Lcal\otimes\Ical_{D}) \ar[r] &0
     }$$
    }
Furthermore, let $\Gamma \subset H^0(X,\Lcal)$ be a submodule such that
    $$
    \Gamma \cap   H^0(X,\Lcal\otimes\Ical_{D}) = (0)
    $$
then, $f$ will be injective and a generator matrix of the code $\C(\Gamma,D)$ (Definition~\ref{def:definitionGoppa}) can be obtained from the evaluation map
    $$
    \Gamma \,\hookrightarrow \, {\F_q}[z]^n
    $$

The length $n$, dimension $k$, memory $m$ and the degree  $\delta$ of the convolutional Goppa codes defined in this section are bounded by
    {\small $$
    n\leq q^4,\quad
    k\leq  h^0(\Pmat^2_{\F_q},\Ocal(r))=\frac{(r+1)(r+2)}2,\quad
    m\leq r, \quad
    \delta\leq\sum_{i=0}^r (i+1)i =\frac13 r(r+1)(r+2)
    $$}

For illustrating this setup and exploring its possibilities, some examples will be provided in the following section.

\section{Explicit Examples}\label{sec:examples}

We will write down explicit examples of the above given construction of  convolutional codes. We will vary $\Gamma$ and $D$  so that codes of different lengths and dimensions will be obtained.

For the sake of clarity, all the examples will deal with a particular case of the situation exposed in \S\ref{sec:CodesP2General}. More precisely, let us take $q=8$, $\F_8$ as base field and $a$ a primitive element such that $a^3+a^2 +1=0$.

Recall that $X:=\Pmat^2_{\F_8[z]} = \Pmat^2_{\F_8}\times \Amat^1$ and that $x_0,x_1,x_2$ denote the homogeneous coordinates in $\Pmat^2$.

Set $\Lcal =    \Ocal_{\Pmat^2_{\F_8[z]}}(\pi_1^* H_\infty)^{\otimes 2}$, then
    $$
    H^0(X,\Lcal) =
    H^0(\Pmat^2_{\F_8[z]},  \Ocal_{\Pmat^2_{\F_8[z]}}(\pi_1^* H_\infty)^{\otimes 2})=
    <1,t,s,t^2, t s ,s^2>
    $$
where $t=\frac{x_1}{x_0}$, $s=\frac{x_2}{x_0}$ are affine coordinates.

In the following examples will study the codes corresponding to certain choices of $D$ and $\Gamma \subseteq  H^0(X,\Lcal)$.

\subsection{Rate $1/3$ codes}

In this case we consider the restriction of the evaluation map to a submodule
$\Gamma\subset H^0(X,\Lcal)$ generated by one section. Here we do not need to care about the properties of $D$ and $\Lcal$ in order to determine the
kernel of the evaluation map; whenever the restriction of the evaluation map to $\Gamma$ is non-zero, $f$ is injective.

Let us consider the 1-dimensional convolutional Goppa code
$\C(D,\Gamma,\psi)$, where $\Gamma\subset H^0(X,\Lcal)$ is the
submodule generated by the section $t+s^2$, and $D$ is consists of
the points
    \begin{equation}\label{eq:3points}
    p_i(z) \,:=\, (a^{2^i} + a^{2^{i-1}} z, a^{2^{i+1}} + a^{2^{i}} z, z)
    \qquad i=1,2,3
    \end{equation}
Then, the map $f$ have the following expression
    $$
    G\,=\,
    \begin{pmatrix}
    a^6 +a z +a^4z^2 &
    a^5 + a^2 z +a z^2 &
    a^3 +a^4 z + a^2 z^2
    \end{pmatrix}
    $$
which is a generator matrix of the code. A straightforward check shows that this generator matrix is canonical.

A control matrix is
    $$
    \begin{pmatrix}
    a^5+a^2z + a z^2  & a^6+ a z+a^4z^2 & 0
    \\
    a^3+ a^4z+ a^2z^2 & 0 & a^6+ a z+a^4z^2
    \end{pmatrix}
    $$

This code has length $3$, dimension $1$, memory $2$,  degree $2$ and free
distance $9$. Further, it attains the generalized Singleton bound (\ref{eq:SingletonBound}) and is, thus, a MDS code.

\subsection{Rate $2/3$ codes}

Continuing with the idea of the previous example, let us consider the submodule
$\Gamma\subset H^0(X,\Lcal)$ generated by the sections $\{t,s^2\}$ and the closed subscheme $D$ given by the three section of equation~(\ref{eq:3points}). It will be seen that the restriction of the evaluation map to $\Gamma$ is injective.

Then, the matrix associated to the restriction of the evaluation map is
    $$
    G\,=\,
    \begin{pmatrix}
    a^2 + a z & a^4 + a^2 z & a+ a^4z
    \\
    a+a^4z^2 & a^2 +a z^2 & a^4+ a^2 z^2
    \end{pmatrix}
    $$
and, since it is injective, it gives us a generator matrix of the code, which is canonical too.

A control matrix is given by the matrix
    $$
    \begin{pmatrix}
    a^4+a z^2 +a^2z^3 &
    a+ a^2z^2 + a^4z^3 &
    a^2 +a^4z^2 +a z^3
    \end{pmatrix}
    $$

This code has length $3$, dimension $2$, memory $2$, degree $3$ and free distance $6$ and it is, thus, a MDS code.

\subsection{Rate $1/4$ codes}

In order to increase the length of the code, four points will be considered. Indeed, $D$ will be now the union of the following four points
    \begin{equation}\label{eq:4points}
    p_i(z) \,:=\, (a^{i} + a^{3 i} z, a^{2 i} +  z, z) \qquad {i=1,\ldots,4}
    \end{equation}
and $\Gamma$ the submodule generated by $t+s^2$.

Then, the restriction of the evaluation map to $\Gamma$ yields a generator matrix of the code $\C(D,\Gamma,\psi)$
    $$
    G\,=\,
    \begin{pmatrix}
    a^3+a^3z +z^2 &
    a^6 +a^6z + z^2 &
    a^6+ a^2 z+z^2 &
    a^5+a^5z +z^2
    \end{pmatrix}
    $$
and a  control matrix is
    $$
    \begin{pmatrix}
    a^6+a^6z +z^2 &
    a^3+a^3z+z^2 & 0 & 0
    \\
    a^6+a^2 z+z^2 &0 &
    a^3+a^3z+z^2 &0
    \\
    a^5+a^5z +z^2 & 0 &0 & a^3+a^3z+z^2
    \end{pmatrix}
    $$

This code has length $4$, dimension $1$, memory $2$, degree $2$ and free distance $12$ and it is, thus, a MDS code.

\subsection{Rate $2/4$ codes}

Finally, let us choose another submodule $\Gamma$ and preserve the same $D$ as above (equation~(\ref{eq:4points})) in order to get a code of length $4$ and dimension $2$. Let  $\Gamma$ the submodule generated by $\{t,s^2\}$.

Then, the restriction of the evaluation map to $\Gamma$ yields the following generator matrix of the code $\C(D,\Gamma,\psi)$
    $$
    G\,=\, \begin{pmatrix}
    a+a^3 z &
    a^2+a^6 z &
    a^3+a^2 z&
    a^4 +a^5 z
    \\
    a^4+z^2 &
    a+z^2 &
    a^5+z^2&
    a^2+z^2
    \end{pmatrix}
    $$
and a control matrix is
    {\small $$
    \begin{pmatrix}
    a^6+a z+z^2+a z^3 &
    a^4+a^2z+a^4z^2+z^3 &
    a+a z+a^6z^2+a^5z^3 &
    0 &
    \\
    a^2+a^2 z+a^5z^2 +a^3z^3 &
    a^4+a^4z+a^3z^2+a^6z^3 &
    0&
    a +a z+a^6z^2+a^5z^3
    \end{pmatrix}
    $$}

This code has length $4$, dimension $2$, memory $2$, degree $3$ and free
distance $8$ and it is, thus, a MDS code.

\section{Conclusions and future work}

The present work is embedded in the study of convolutional codes
from an algebraic geometric point of view. It is in particular a
further step in the construction of convolutional codes by means of
algebraic geometric tools. The main contribution with respect to
previous works is the generalization of the setting where codes are
defined. We present in this paper the construction of convolutional
codes on higher dimensional varieties using a rather small alphabet. In particular, the
construction is detailed for the variety
$\Pmat^2_{\F_q}\times\Amat^1$. Although it is in general a hard task
to compute the parameters of the codes obtained we are able to find
different MDS convolutional codes, and to illustrate this fact
examples of codes with different rates are also presented for the case of the projective plane over $_{\F_8}$.

The results obtained confirm our belief that this is a promising
research line. Future steps could be to determine conditions on the
geometric elements to obtain good codes or to adapt decoding
algorithms to decode them.

\bibliographystyle{amsplain}



\end{document}